\documentclass{mem}
\usepackage{natbib}\usepackage{txfonts}\usepackage{balance}
\usepackage{graphicx}
\usepackage[a4paper,breaklinks,dvipdfm]{hyperref}
\idline{999}{999}
\begin{document}

\title{The Asymptotic Giant Branches of GCs:\\ Selective Entry Only}
\subtitle{}

\author{
S. W.\, Campbell\inst{1},
V.\, D'Orazi\inst{2,1},
T. N.\, Constantino\inst{1}, 
D.\, Yong\inst{3},
J. C.\, Lattanzio\inst{1},
G. C.\, Angelou\inst{1},
E. C.\, Wylie-de Boer\inst{3},
R. J.\, Stancliffe\inst{4,3},
S. L.\, Martell\inst{5},
\and
F.\, Grundahl\inst{6}
        }

\offprints{S.W. Campbell}

\institute{
           Monash Centre for Astrophysics (MoCA), 
           Monash University, Building 28, Victoria, Australia 3800.
\and
           Department of Physics \& Astronomy, Macquarie University,
           Balaclava Rd, 
           North Ryde, Sydney, NSW, Australia 2019.
\and
           Research School of Astronomy and Astrophysics, 
           Australian National University, 
           Weston, ACT 2611, Australia.
\and       
           Argelander-Insitut f\"{u}r Astronomie, Universit\"{a}t Bonn, Auf dem
           H\"{u}gel 71, 53121 Bonn, Germany.
\and
           Australian Astronomical Observatory, North Ryde, NSW 2113,
           Australia.
\and
           Department of Physics and Astronomy, 
           Aarhus University, Ny Munkegade, 8000 Aarhus C, Denmark.
}

\authorrunning{Campbell et al.}

\titlerunning{AGBs in GCs}

\abstract{The handful of available observations of AGB stars in Galactic
  Globular Clusters suggest that the GC AGB populations are dominated by
  cyanogen-weak stars. This contrasts strongly with the distributions in
  the RGB (and other) populations, which generally show a 50:50 bimodality
  in CN band strength. If it is true that the AGB populations show very
  different distributions then it presents a serious problem for low mass
  stellar evolution theory, since such a surface abundance change going
  from the RGB to AGB is not predicted by stellar models. However this is
  only a tentative conclusion, since it is based on very small AGB sample
  sizes. To test whether this problem really exists we have carried out an
  observational campaign specifically targeting AGB stars in GCs. We have
  obtained medium resolution spectra for about 250 AGB stars across 9
  Galactic GCs using the multi-object spectrograph on the AAT
  (2df/AAOmega). We present some of the preliminary findings of the study
  for the second parameter trio of GCs: NGC 288, NGC 362 and NGC 1851. The
  results indeed show that there is a deficiency of stars with strong CN
  bands on the AGB. To confirm that this phenomenon is robust and not just
  confined to CN band strengths and their vagaries, we have made
  observations using FLAMES/VLT to measure elemental abundances for NGC
  6752. We present some initial results from this study also. Our sodium
  abundance results show conclusively that only a subset of stars in GCs
  experience the AGB phase of evolution. This is the first direct, concrete
  confirmation of the phenomenon.

\keywords{AGB stars --
          Globular cluster --
          Abundances -- 
          Cyanogen
          }
          } 

\maketitle{}
%
%
\section{Introduction}
One of the first chemical inhomogeneities discovered in Galactic globular
clusters (GCs) was that of the molecule cyanogen (CN, often used as a proxy
for nitrogen; eg. \citealt{NCF81}). The CN band strengths show strong
star-to-star variations, with the populations falling into two (or more)
populations: CN-weak stars and CN-strong stars. This CN bimodality is seen
in all evolutionary phases studied in detail thus far. 

Due to the paucity of asymptotic giant branch (AGB) stars in GCs (a result
of their short lifetimes), plus the difficulty in splitting the red giant
branch (RGB) and AGB in colour-magnitude diagrams (CMDs), there have been
very few systematic observational studies of the CN anomaly in the AGB
populations of globular clusters (\citealt{M78} is one of which we are
aware).  What little that has been done has been an aside in more general
papers (e.g. \citealt{NCF81}, \citealt{BSH93}, \citealt{ISK99},
\citealt{SIK00}). However these studies have hinted at a tantalising
characteristic: most (observed) GCs show a lack of CN-strong stars on the
AGB. If this is true then it is in stark contrast to the RGB and earlier
phases of evolution, where the ratio of CN-Strong to CN-Weak stars is
roughly 50:50 in many clusters. It is also strange from a stellar evolution
theory perspective, since a surface abundance change going from the RGB to
AGB is not predicted by standard models. The possible existence of this
phenomenon is however based on studies with small sample sizes.

Here we present some preliminary results of from our large study that
increases the GC AGB sample sizes substantially. With this new information
we hope to confirm or disprove the existence of the abundance differences
between the AGB and other phases of evolution.
%
%
\section{Observations: CN Band Strengths}
A vital ingredient in being able to find significant numbers of AGB stars
in globular clusters is having photometry good enough to separate the AGB
from the RGB. For the current study we have used the \citet{walker92}
sample for NGC 8151, plus a range of other CMDs, mostly from \cite{GCL99}.
In total our sample consists of a database of $\sim 800$ stars at various
stages of evolution (RGB, HB \& AGB), in 10 GCs.

The medium-resolution observing run consisted of 5 nights on the AAT. We
used the multi-object spectroscope, AAOmega/2dF \cite{smith04}. On the blue
arm we used the 1700B grating, which gave a spectral coverage of 3755 to
4437 $\AA$, which includes the violet CN bands. The resolution of the
spectra is $\sim 1.2 \AA$.  Near the CN bands the S/N $\gtrsim 20$.

To quantify the CN band strengths in each star we use the S(3839) CN index
of \cite{NCF81}. We then remove the trend with temperature (see
\citealt{NCF81,campbell12}), finally giving the $\delta$S(3839) value.
%
%
%
\section{Intriguing Results: CN Distributions and HB Morphology}
In this conference proceedings we present preliminary results for the
second parameter trio NGC 288, NGC 362 and NGC 1851. These GCs all have
similar metallicities ([Fe/H] $\sim -1.3$) but different HB morphologies
(blue, red and blue+red HBs respectively; \citealt{bellazzini01}).

%
\begin{figure}[ht]
\centering
\resizebox{0.8\hsize}{!}{\includegraphics[clip=false]{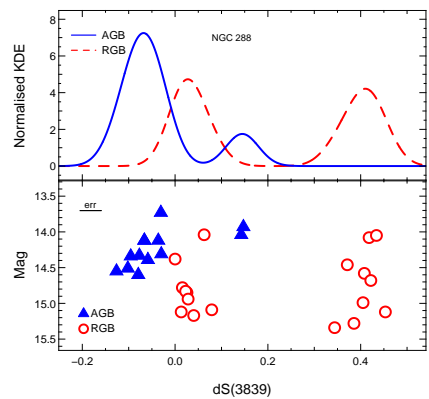}}
\caption{\footnotesize Preliminary CN results for NGC 288 from
  AAT/2dF. Lower panel shows the $\delta$S(3839) results for each of the
  stars in our sample. The upper panel shows the same data represented by a
  kernel density estimate histogram (kernel bandwidth $= 0.035$).}
\label{288results}
\end{figure}
\begin{figure}[ht]
\centering
\resizebox{0.8\hsize}{!}{\includegraphics[clip=false]{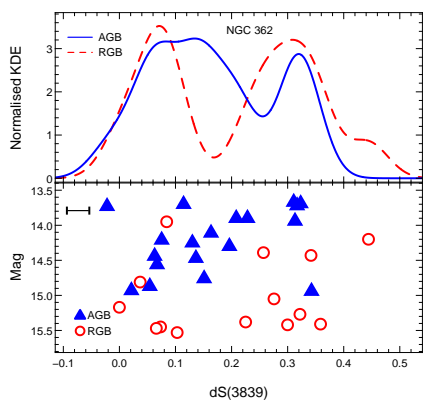}}
\caption{\footnotesize Same as Fig. \ref{288results} but for NGC 362.}
\label{362results}
\end{figure}
\begin{figure}[ht]
\centering
\resizebox{0.8\hsize}{!}{\includegraphics{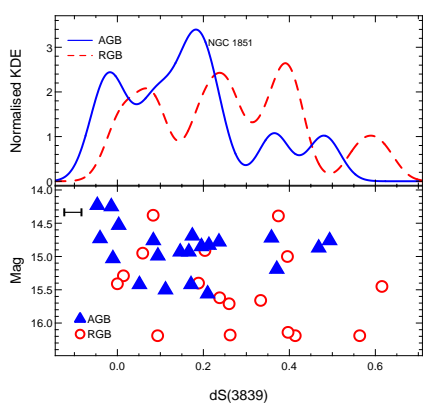}}
\caption{\footnotesize Same as Fig. \ref{288results} but for NGC 1851.}
\label{1851results}
\end{figure}
%
In Figure \ref{288results} we show the $\delta$S(3839) results against
magnitude for NGC 288. This GC shows the classic bimodality in CN on the
RGB. However the AGB is clearly dominated by CN-weak stars. This is a very
large change in CN population proportions, especially considering that all
other phases of evolution show similar ratios to the RGB. Interestingly
there is a hint of bimodality on the AGB also, with two stars being
significantly more CN-strong than the other AGB stars (although still
considered CN-weak compared to the RGB stars). In Figure \ref{362results}
we show the $\delta$S(3839) results against magnitude for NGC 362. Again
the classic bimodality can be seen in the RGB. However the AGB results are
less clear. We suggest that the AGB CN distribution is consistent with
there being no change in proportions between the RGB and AGB. A comparison
of the NGC 288 and NGC 362 results suggest that the abundance anomalies are
related to HB morphology, such that GCs with blue HBs show a lack of
CN-strong stars on the AGB whilst GCs with red HBs do not. Finally, in
Figure \ref{1851results} we show the $\delta$S(3839) results against
magnitude for NGC 1851. This GC has a combination of a red and a blue HB,
so could be considered an intermediate case between NGC 288 and NGC
362. The RGB results are quite different to the other GCs -- the
distribution appears to be quadrimodal. This is lent more weight by the AGB
distribution, which is also quadrimodal, although the majority of the stars
are CN-weak. We have discussed this interesting case elsewhere in the
context of a GC merger scenario (\citealt{campbell12}).
\section{Conclusive Proof: High-resolution Elemental Abundances}
\begin{figure}[ht]
\centering
\resizebox{0.9\hsize}{!}{\includegraphics{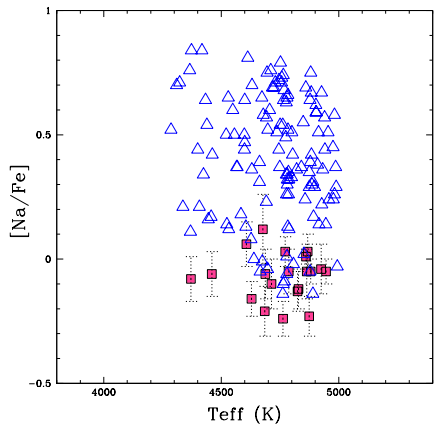}}
\caption{\footnotesize Preliminary Na results from VLT/FLAMES spectra for
  our NGC 6752 AGB sample (red squares with error bars). Sodium for RGB
  stars from \cite{carretta07} are shown for comparison (open triangles).}
\label{6752Naresults}
\end{figure}

The preliminary results of our CN study strongly support the unexpected
phenomenon in which CN-strong stars seem to `disappear' between the RGB and
AGB, leaving CN-weak dominated AGBs. However there is the possibility that
the measurements of CN in AGB stars are biased in some way, either by
gravity or temperature differences as compared to the RGB stars. Cyanogen
molecular band strengths are also affected by the distribution of C, N and
O. For these reasons the CN results are not considered 100\% robust. In
order to conclusively determine the proportions of polluted and
non-polluted stars on the AGB we have undertaken a high resolution study of
NGC 6752 using VLT/FLAMES. In \cite{campbell10} we reported our preliminary
CN results for NGC 6752. This cluster is similar to NGC 288
(Fig. \ref{288results}), showing a clear bimodality in the RGB sample and a
CN-weak dominated AGB. In Figure \ref{6752Naresults} we show our
preliminary VLT/FLAMES results for sodium in NGC 6752 AGB stars.  A
striking result can be seen -- \textit{every single AGB star is
  Na-poor}. This compares with a ratio of roughly 60:40 CN-strong to
CN-weak in the RGB and other populations. It now appears certain that
second-generation (N and Na-rich, C and O-poor stars) do not evolve to the
AGB phase in NGC 6752, and probably many other GCs.

One explanation for this phenomenon is that the two populations in NGC 6752
have different He abundances in addition to their C, N, O and Na abundance
differences (\citealt{NCF81,dantona02}).  The He-rich material would also
be N-rich due to CNO cycling. The He-rich stars would then evolve to
populate the bluest end of the HB -- and not ascend the AGB -- leaving only
CN-weak stars to evolve to the AGB. This ties in well with the recent
findings of \cite{villanova09} and \cite{marino11} (in NGC 6752 and M4
respectively) that the Na-rich stars populate only the blue ends of the
HBs, whilst the Na-poor stars populate the red(der) ends of the HBs.

The Na results for NGC 6752 presented here represent the first conclusive
proof that only certain stars make it to the AGB phase of evolution
-- the other stars must go directly to the white dwarf phase. 
%
%
%
\begin{acknowledgements}
Thanks to the LOC \& SOC of of the conference held at Rome Observatory.
\end{acknowledgements}
\bibliographystyle{aa}

\end{document}